\documentclass{article}
\usepackage{graphicx}
\usepackage{geometry}
\begin{document}

\title{An Overview of a Grid Architecture for Scientific Computing}

\author{A.~W\"{a}\"{a}n\"{a}nen$^a$, M.~Ellert$^b$, A.~Konstantinov$^c$, B.~K{\'o}nya$^d$, O.~Smirnova$^d$}
\date{{\footnotesize 
$^a$Niels Bohr Institutet for Astronomi, Fysik og
  Geofysik,\\ Blegdamsvej 17, DK-2100, Copenhagen \O, Denmark\\
$^b$Department of Radiation Sciences, Uppsala University,\\
  Box~535, 751 21 Uppsala, Sweden\\
$^c$University of Oslo, Department of Physics,\\
  P.~O.~Box~1048, Blindern, 0316 Oslo, Norway\\
$^d$Particle Physics, Institute of Physics, Lund University,\\
  Box~118, 22100 Lund, Sweden}}

\maketitle

\begin{abstract}
  This document gives an overview of a Grid testbed architecture proposal for
  the NorduGrid project. The aim of the project is to
  establish an inter-Nordic testbed facility for
  implementation of wide area computing and data handling. The architecture
  is supposed to define a Grid system suitable for solving data intensive
  problems at the Large Hadron Collider at CERN. We present the
  various architecture components needed for such a system. After that we go
  on to give a description of the dynamics by showing the task flow.
\end{abstract}

\section{Introduction}

This document assumes basic knowledge of the computing {\em Grid}
concept, which is a paradigm for the modern distributed computing and
data handling.  For a general introduction to Grid computing the
reader is referred to eg.~\cite{grid-blueprint}. The most common
starting point for constructing a computing Grid is the Globus
Toolkit\footnote{Globus Project and Globus Toolkit are trademarks
held by the University of Chicago.}~\cite{globus-toolkit}.  This
toolkit provides a Grid API and developing libraries as well as basic
Grid service implementations.

The NorduGrid~\cite{NorduGrid} project is a common effort by the
Nordic\footnote{The term {\em Nordic} covers the countries:
Denmark, Norway, Sweden and Finland.} countries to create a Grid
infrastructure, making use of the available middleware. Through the
European DataGrid project (EDG)~\cite{EDG} the NorduGrid project has
had extensive experience with the Globus Toolkit and with deploying
and using a Grid Testbed. During this we have found some shortcomings
in the Globus Toolkit and some problems with the EDG Testbed
architecture that we would like to address on a Grid testbed in the
Nordic countries. In this paper we present a proposal for a Grid
architecture for a production testbed at the LHC~\cite{LHC}
experiments. It is not the intent to define a general Grid system, but
rather a system specific for batch processing suitable for problems
encountered in High Energy Physics. Interactive and parallel
applications have not been considered. It is our goal to deploy a Grid
testbed based on this architecture for the LHC Data challenges in the
Nordic countries.

We will start by describing the various components of the system. Focus will
be given to the components which we have developed ourselves or taken from
others, but heavily modified. After the description we will give a review of
the task flow and the communication between the various components.

It is assumed that the reader is familiar with the major components of the
Globus Toolkit since the NorduGrid Testbed architecture uses this as the
foundation. An example is the underlying security model - the Grid
Security Infrastructure (GSI) - which is a based on the {\texttt X.509}
certificate system and takes care of authentication and authorisation and the
GRIS and GIIS of the information system (for an explanation of the GRIS,
GIIS, VO terms see~\cite{MDS} and~\cite{VO}).

One of the aims of the architecture was to make the system able to install on
top of an existing Globus installation. Rather than modifying the Globus
Toolkit, we have made a clear boundary between the two architectures, thus
enabling them to coexist and try out this new system without destroying an
already working Globus installation.

\section{The NorduGrid Architecture Components}

\subsection{Computing Element - CE}

The {\em computing element} (CE) is the backend of the system. It is where
the execution of the application is performed. This is a very general element
which can be anything from a single processor to complex computing clusters.
In our case it is at present limited to simple PBS (Portable Batch System)
clusters.

One of our basic ideas about Grid implementations is that it should not
impose any restriction on the local site. Therefore it should be possible to
use existing computing clusters and place them on the Grid with minor or no
special reconfiguration. One thing that we have stressed, which should not be
imposed on the site, is that the individual worker nodes (WN's) of the
cluster can not be required to have direct access to the Internet. This
excludes dependence on eg. global filesystems like AFS and direct download
from the Internet from any WN. Grid services will thus only be run from a
front-end machine. This scenario does not however exclude the use of local
network filesystems and NFS is actually often used for cluster-wide
application distribution.

\subsection{Storage Element - SE}

Similar to the computing element, the {\em storage element} (SE) is the
common term for another of the basic Grid resources - storage. The storage
can be as simple as a standard filesystem or eg. a database. The
authorisation is handled using Grid credentials (certificates). A SE can be
{\em local} or {\em remote} to a CE. Here {\em local} means that access is
done via standard filesystem calls (eg. open, read and close) and is usually
realised by a NFS server. A remote SE is usually a stand-alone machine
running eg.  GridFTP~\cite{GridFTP} server with local file storage. Data
replication is done by services running on the SE.

A dedicated pluggable GridFTP server 
is beeing
developed for use on the SE. At the moment a simple file access plugin
exists. The main reason for this is to have a way to provide a
consistent certificate-based data access to the data.  At least one
other Grid solution to a certificate-based filesystem
exists~\cite{slashgrid}. One advantage of the GridFTP approach is that
it is done entirely in user space and thus is very portable.

\subsection{Replica Catalog - RC}

The information about replicated data is contained in the
{\em~Replica~Catalog} (RC). This is an entirely add-on component to the
system and as such is not a requirement.

\subsection{Information System - IS}

A stable, robust, scalable and reliable information system is the cornerstone
of any kind of Grid system. Without a properly working information system it
is not possible to construct a functional Grid. The Globus Project has laid
down the foundation of a Grid information system with their LDAP-based
Metacomputing Directory Service (MDS)~\cite{MDS}. The NorduGrid information
system is built upon the Globus MDS.

The information system described below forms an integral part of the
NorduGrid Testbed Architecture. In our Testbed, the NorduGrid MDS plays a
central role: all the information related tasks, like resource-discovery,
Grid-monitoring, authorised user information, job status monitoring, are
exclusively implemented on top of the MDS. This has the advantage that all
the Grid information is provided through a uniform interface in an inherently
scalable and distributed way due to the Globus MDS. Moreover, it is sufficient
to run a single MDS service per resource in order to build the entire
system. In the NorduGrid Testbed a resource does not need to run dozens of
different (often centralized) services speaking different protocols: the
NorduGrid Information System is purely Globus MDS built using only the LDAP
protocol.

The design of a Grid information system is always deals with questions like
how to represent the Grid resources (or services), what kind of information
should be there, what is the best structure of presenting this information to
the Grid users and to the Grid agents (i.e. Brokers).  These questions have
their technical answers in the so-called LDAP schema files. The Globus
Project provides an information model together with the Globus MDS. We found
their model unsuitable for representing computing clusters, since the Globus
schema is rather single machine oriented. The EDG suggested a different CE
model which we have evaluated~\cite{CE-evaluation}.  The EDG's CE-based
schema fits better for computing clusters. However, its practical usability
was found to be questionable due to improper implementation.

Because of the lack of a suitable schema, NorduGrid decided to design its own
information model (schema) in order to properly represent and serve its
Testbed. A working information system, as part of the NorduGrid Architecture,
has been built around the schema. Nevertheless, NorduGrid hopes that in the
not-so-far future a usable common Grid information model will emerge. We
think that the experience of the NorduGrid users gained with our working
system will provide a useful feedback for this process.

\subsubsection{The Information Model.}
The NorduGrid Testbed consists of different resources (they can be referred
as services), located at different sites. The list of implemented Grid
services contains computing resources (Linux clusters operated by PBS),
SEs (at the moment basically disk space with a GridFTP server) and
RCs. The designed information model naturally maps these
resources onto a LDAP-based MDS tree, where each resource is represented by
an MDS entry. In this tree, each NorduGrid resource operates as a separate
GRIS. The various resources (GRIS's) can be grouped together to form Virtual
Organisations (VO) which are served by a GIIS (i.e. in our present Testbed
configuration, the resources within a country are grouped together to form a
VO). The structure created this way is called a {\em hierarchical MDS tree}.

\begin{figure}[ht]
  \centering
\includegraphics[width=12cm]{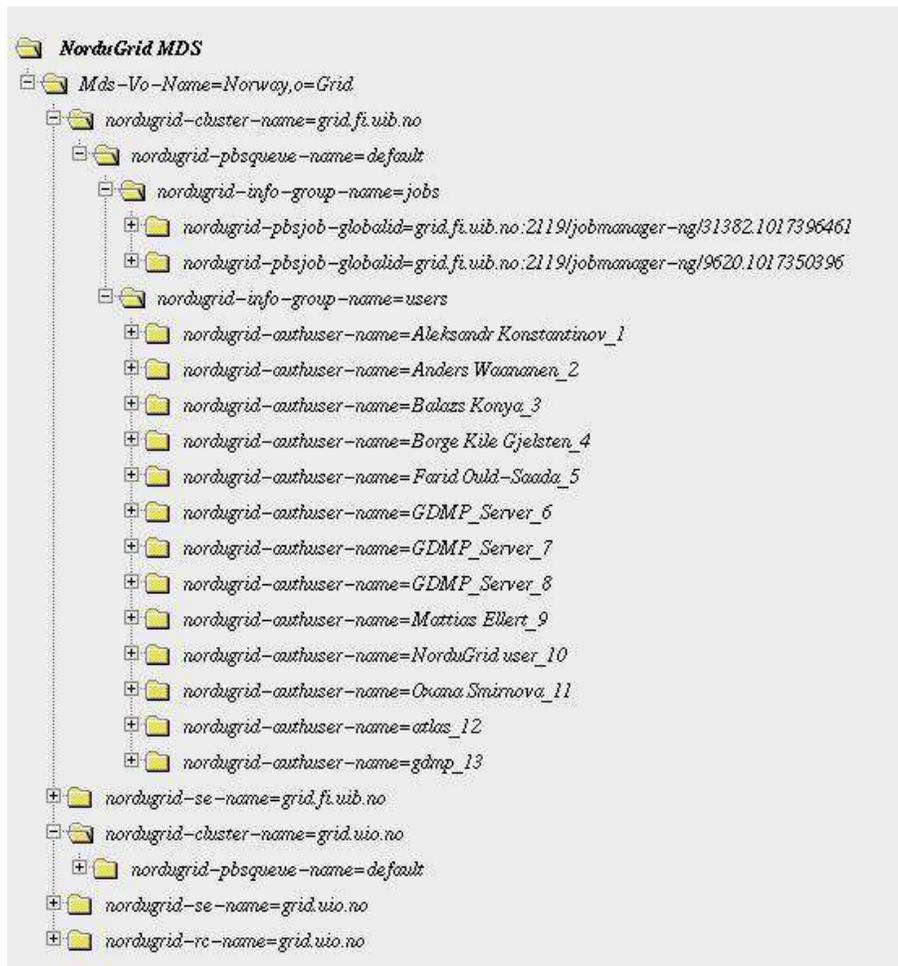}
\caption{\label{fig:MDStree} The Norway branch of the NorduGrid MDS tree}
\end{figure}

\begin{figure}[ht]
  \centering
  \includegraphics[width=12cm]{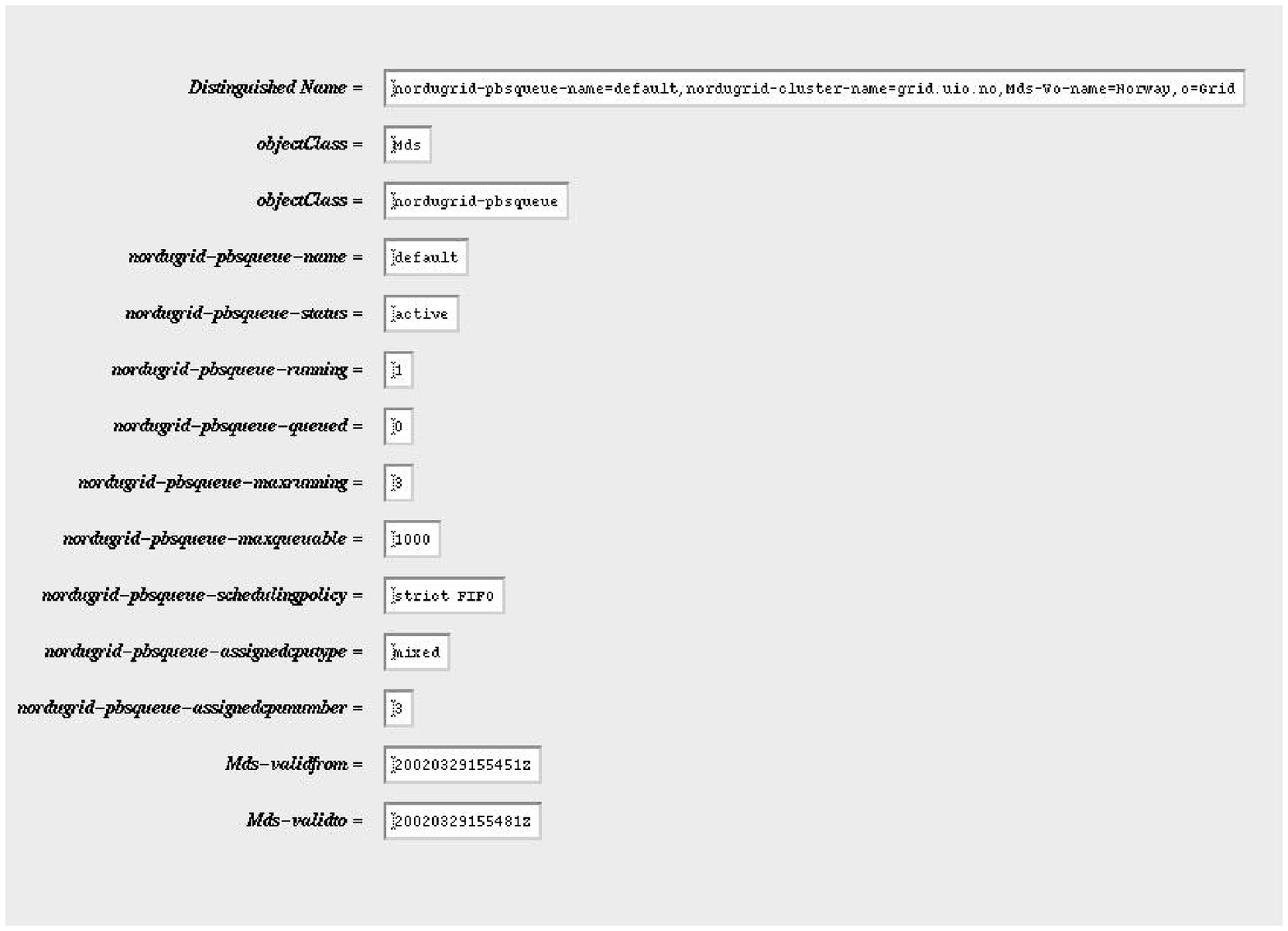}
  \caption{Example nordugrid-pbsqueue entry}
  \label{fig:pbsqueue}
\end{figure}

\begin{figure}[ht]
  \centering
  \includegraphics[width=12cm]{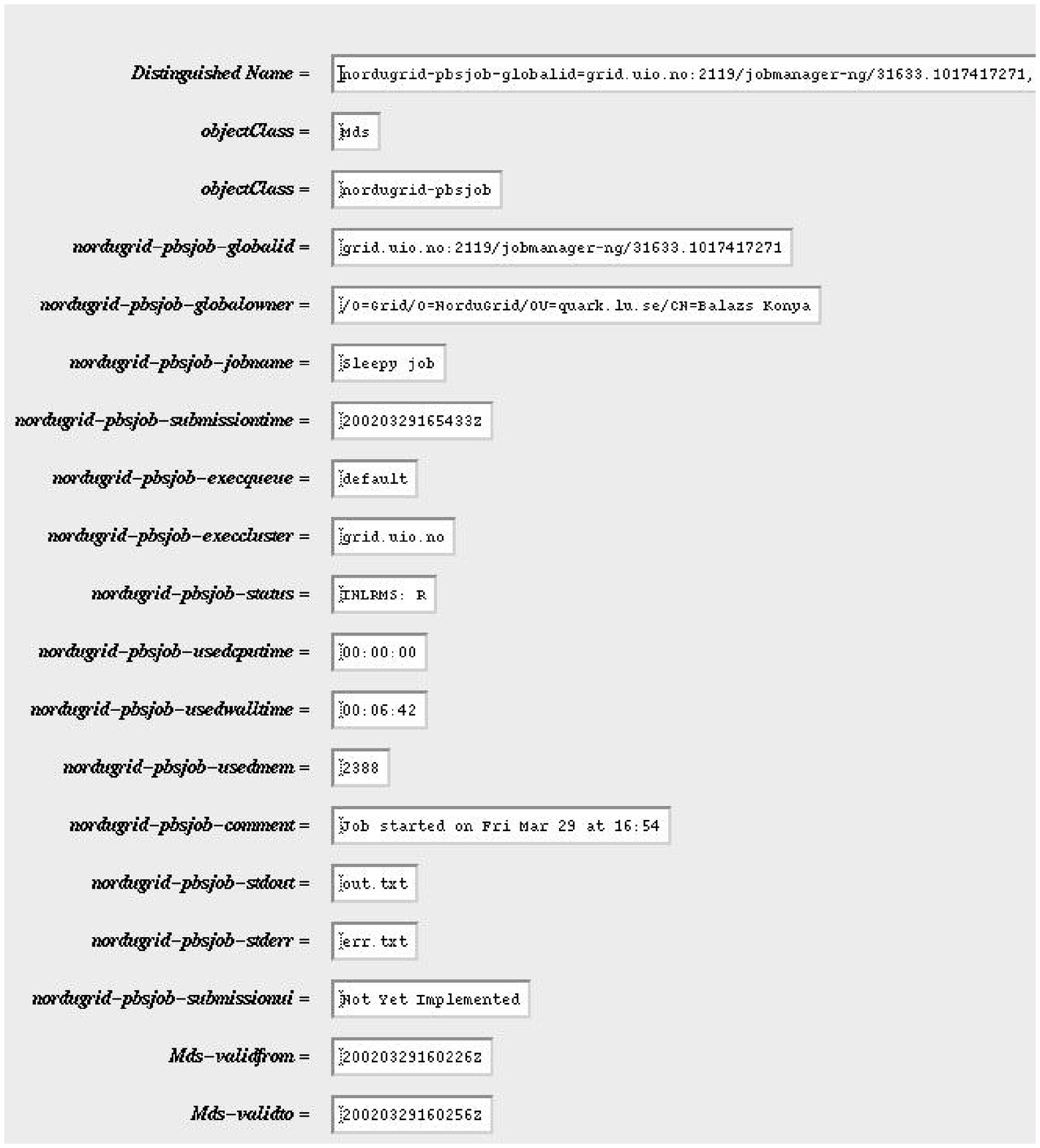}
  \caption{Example nordugrid-pbsjob entry}
  \label{fig:pbsjob}
\end{figure}

The NorduGrid schema is a true mirror of our architecture: it contains
information about computing clusters ({\tt nordugrid-cluster} objectclass),
storage elements ({\tt nordugrid-se} objectclass) and replica catalogs ({\tt
  nordugrid-rc} objectclass). In Fig.~\ref{fig:MDStree}, a part of the
NorduGrid MDS tree~\cite{ldapexplorer} is shown.  The clusters provide access
to different PBS queues which are described by the {\tt nordugrid-pbsqueue}
objectclass (Fig.~\ref{fig:pbsqueue} shows an example queue entry).  Under
the queue entries, the nordugrid-authuser and the nordugrid-pbsjob entries can
be found grouped in two distinct sub-trees (the branching is accomplished with
the {\tt nordugrid-info-group} objectclass). The {\tt nordugrid-authuser}
entry contains all the user-dependent information of a specific authorised
Grid user. Within the user entries, the Grid users can find, among other
things, how many CPUs are available for them in that specific queue, what is
the disk space a job can consume, what is an effective queue length
(taking into account the local UNIX mappings), etc. The {\tt
  nordugrid-pbsjob} entries (see Fig.~\ref{fig:pbsjob} an example) describe
the Grid jobs submitted to a queue. The detailed job information includes
the job's unique Grid ID, the certificate subject of the job's owner and the
job status.
  
The NorduGrid information system has been designed to be able to effectively
serve the User Interface (UI) (job status query commands, free resource
discovery utilities), the brokering agent (in the present implementation it
is integrated with the UI's job submission command) and a general Grid user
who can access the Grid information either directly with a simple
{\tt ldapsearch} or through an LDAP-enabled Grid portal.

\subsection{Grid Manager - GM}

In our model, job management is handled by a single entity which we call the
{\em~Grid~Manager} (GM). It is the job of the GM to process user requests and
prepare them for execution on the CE. It also takes care of post-processing
the jobs before they leave the CE. In the Globus Toolkit context, the GM
takes care of what is normally done by the Globus job-manager
 backend scripts.
In fact it is
installed in a similar way to a standard Globus jobmanager and can work
perfectly together with already existing jobmanagers. Authorisation and
authentication is still done by the Globus gatekeeper.

The status of each job is recorded in a special status directory which also
contains
informational
files needed by the GM.

\subsubsection{Grid Manager tasks.}
The Grid Manager uses a job ID assigned by the Globus
gatekeeper/jobmanager to distinguish between job requests. 
It starts by
creating a session directory reachable by the WN and parsing the job
request. The job request is passed in a resource specification language. This
specifies, among other things, the list of input files needed for the job
execution. 
The Grid Manager then proceeds to download files to the session directory
from remote SEs or copy files from 
local SEs as specified by the job request.Files can also be uploaded
by the user submitting the job. Once all files are in place the GM
submits the job to the local scheduling system (PBS). All information
about job status must be retrieved through the MDS.  Communication to
the job such as cancellation is handled through the GM 
as a specific RSL requests.
When the job leaves the WN(s) it is the responsibility of
the GM to clean up afterwards. The GM also manages the output files
and uploads them to local or remote storage elements and registers
uploaded files to the Replica Catalog.

\subsection{User Interface (UI)}

In contrast to eg. the Grid implementation in the EDG, the NorduGrid
{\em~User~Interface~(UI)} has significantly more responsibility. This
is mainly due to the choice of having the resource broker placed at
the UI rather than having it as a central service. We found that the
EDG implementation with a central broker, where all job requests as
well as data payload had to pass through, would be a single point of
failure and non-scalable.

The NorduGrid UI is at present command-line driven, while a web based
solution is foreseen in the future. The UI is responsible for
generating the user request in a {\em Resource Specification Language}
(RSL) based on the user input. The RSL we use has additional
attributes to those provided by the Globus Toolkit~\cite{RSL}. This
xRSL has been enhanced to support enriched input/output capabilities
and more specification of PBS requirements. All unneeded Globus
attributes has been deprecated.

The UI matches the request or job options with the available resources
as reported by the MDS, and returns a list of matching resources. Job
options can reflect eg.:

\begin{itemize}
\item required CPU time
\item required system type
\item required disk space
\item required runtime environment (eg. application software)
\item required memory
\item required data from SE's
\end{itemize}


\section{Task flow}

\begin{figure}[ht]
  \centering \par
\includegraphics[width=12cm]{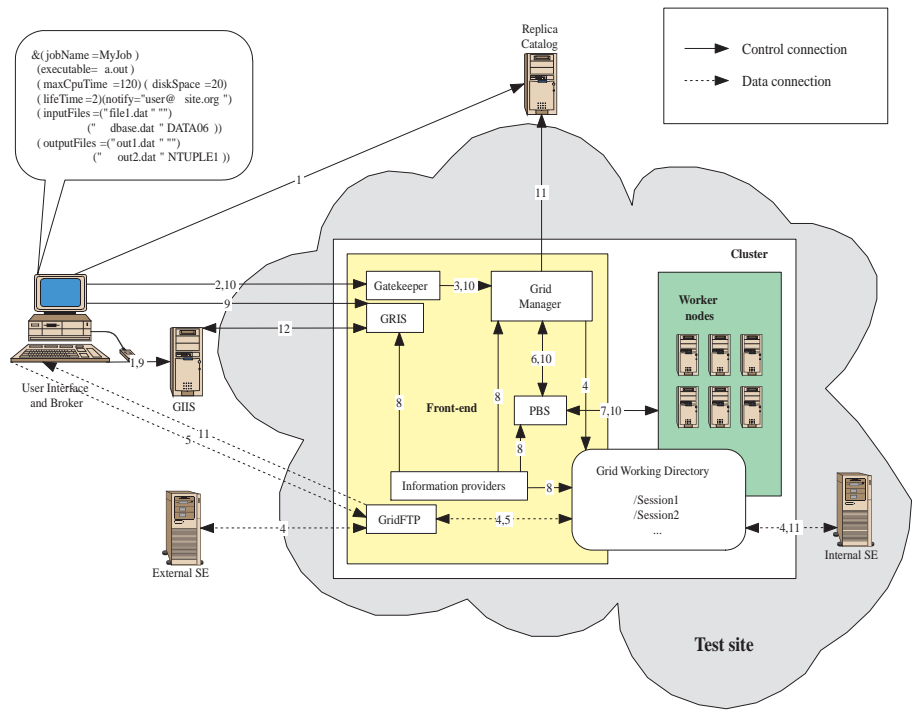}
  \caption{NorduGrid task flow}
  \label{fig:taskflow}
\end{figure}

In this section we describe the life of a job and expose the functions
of the various components of the NorduGrid architecture. An overview
of the system with task flow can be seen in Fig.~\ref{fig:taskflow}.

The numbers on the figure refers to tasks which we describe below.

\begin{enumerate}
\item The User Interface does a filtered query against the GIIS and
  query the Replica Catalog to get the location of input data. Based on
  these responses
  and its brokering algorithm 
  the Broker within the UI selects a remote resource.
\item The UI 
  contacts
  the 
  selected resource by submitting the 
  xRSL file to the Gatekeeper of the resource, along with 
  the (resolved by the Broker) physical file names.
\item The Gatekeeper does the required authentication and authorisation and
  passes the job request to the Grid Manager.
\item Grid Manager creates the session directory and downloads or copies files
  from Storage Elements to this directory
\item User Interface uploads input files and executables via GridFTP
  to the
  session directory of the job
\item After all files are pre-staged, the Grid Manager submits the job to the
  local resource management system (PBS)
\item PBS schedules 
  and executes
  the job on the Worker Nodes in the normal way
\item 
  On request,
  Information Providers collect job
  , user, queue and cluster
  information, disk usage information
  and writes the information to the MDS
\item 
  Job status information is retrieved from the MDS, the
  User Interface monitors the job status by querying the 
  Information System. E-mail notification
  to the user during the various job stages by the Grid Manager is
  also possible.
\item User Interface may cancel
  and clean jobs by sending RSL request with special attribute set 
  through the Gatekeeper to the Grid Manager. The Grid Manager will
  then take care of
  stopping the job and removing the job session directory if requested
\item When the job finishes the Grid Manager moves requested output results
  to Storage Elements
  and does registration in the Replica Catalog as specified by the user.
  The Grid Manager takes care of cleaning up the session directory
  after its lifetime has exceeded, too
  While the directory exists, the User Interface can download the
  specified
  output files produced by the job.
\item GIIS queries GRIS on demand based on cache timeout values in order to
  provide fresh
  enough 
  information. The single GIIS in the figure really represents a
  whole Virtual Organisation hierarchy. In NorduGrid testbed the hierarchy
  has a central NorduGrid GIIS which connects to country level GIIS's. The
  country level GIIS's either connects directly to resources or have an
  institutional layer in between.
\end{enumerate}

In the figure data and control connections are specified. In designing the
system we have been very conscious about the location of data bottlenecks and
places where the system will have scalability problems. The only data
intensive transfers occur between the 
Computing Element
and the Storage
Elements and the User Interface. In contrast to the EDG there is no single
point where all data has to pass through. All transfers are truly
peer-to-peer.

\section{Conclusion and Outlook}

A very early prototype implementation of the architecture exists and is
being tested and further developed. Scalability tests still has to be
performed.

In order to have a functional testbed within a short time-frame, there are
some areas which we have not given much attention and have taken simple but
still secure and functional solutions. In the future we plan to include, for
example a more advanced authorisation and accounting system.

Once we get experience with the system we are going to make the Broker more
sophisticated and make it able to perform user dependent choices such as RC
location and preferred resource selection.

\section{Acknowledgment}

We are very grateful for the developers of the Globus Project for their
willingness to answer questions and providing an open development
environment with access to early version of their toolkit.

As participants in the European DataGrid we have been provided access to a
Testbed and the collaboration has been very fruitful.

The NorduGrid project - {\em A Nordic testbed for wide area computing and
  data handling } ({\texttt http://www.nordugrid.org/}) - is funded by the
Nordunet2 program ({\texttt http://www.nordunet2.org/}.

\end{document}